\documentclass[pre,amsmath,twocolumn,showpacs]{revtex4-1}
\usepackage{bm}
 \usepackage{amssymb}
 \usepackage{graphicx}
 \usepackage{subfigure}
 \usepackage[dvips]{color}

\begin{document}

 \title{Finite-density effects in the Fredrickson-Andersen and Kob-Andersen kinetically-constrained models}

 \author{Eial Teomy}
 \email{eialteom@post.tau.ac.il}
 \author{Yair Shokef}
 \email{shokef@tau.ac.il}

 \affiliation{School of Mechanical Engineering, Tel Aviv University, Tel Aviv 69978, Israel}

 \begin{abstract}

We calculate the corrections to the thermodynamic limit of the critical density for jamming in the Kob-Andersen and Fredrickson-Andersen kinetically-constrained models, and find them to be finite-density corrections, and not finite-size corrections. We do this by introducing a new numerical algorithm, which requires negligible computer memory since contrary to alternative approaches, it generates at each point only the necessary data. The algorithm starts from a single unfrozen site and at each step randomly generates the neighbors of the unfrozen region and checks whether they are frozen or not. Our results correspond to systems of size greater than $10^{7}\times10^{7}$, much larger than any simulated before, and are consistent with the rigorous bounds on the asymptotic corrections. We also find that the average number of sites that seed a critical droplet is greater than $1$.
 
 \end{abstract}


 \maketitle

\section{Introduction}

Increasing the density of particles in granular materials causes them to undergo a transition from a fluid-like state, in which the particles can move relatively freely, to a jammed state, in which almost none of the particles can move~\cite{jamming,hecke}. In glasses, a similar transition occurs when the temperature is decreased \cite{glass,glass2}. The various kinetically-constrained models \cite{review,review2,east,northeast,kronig} capture the essence of the glass or jamming transitions and there has been much recent activity on them; Some of these models simulate the way the particles block each other's movement by saying that a particle can move only if its neighbors satisfy some condition \cite{sellitto,knights,DFOT,spiral,spiral3d,jeng,elmatad}.
Other models add driving forces which simulate the resistance of jammed systems to forces \cite{fieldings,shokef,sellitto2,driving2,driving}.

Two of the most studied kinetically-constrained models are the Fredrickson-Andersen (FA) \cite{fa,fa2} and the Kob-Andersen (KA) \cite{ka} models. In the versions of the models we consider here, the system is coarse-grained to a two-dimensional square lattice, and each site is in one of two states, $0$ or $1$. In the FA model, state $1$ represents a high density region in granular systems and an active region in glasses, while state $0$ represents either a low density region or an inactive region in granular matter and glasses respectively. A site can change its state from $0$ to $1$ and vice versa, with a temperature-dependent rate, if at least $m=2$ of its four nearest neighbors are in state $0$. In the KA model, state $1$ represents an occupied site and state $0$ a vacant site. A particle can move to a vacant nearest neighbor site if at least $m=2$ of its nearest neighbors are vacant before and after the move. These models can be expanded to higher dimensional hyper-cubic lattices with a general number $m$ of vacant neighbors needed for movement \cite{teomy2,balogh}. We restrict ourselves in this article to the $m=2$ models in a two-dimensional square lattice. Our work can be easily extended to higher dimensional models with $m=2$, but extending it to models with $m\geq3$ is much more complicated.

In order to investigate the jamming transition in kinetically-constrained models, one defines an order parameter $n_{F}$ as the average fraction of particles that will never be able to move, and we call these particles \textit{frozen}. It was proven that in the thermodynamic limit, none of the particles are frozen in either the FA model \cite{fa2} or the KA model \cite{toninelli} for any finite density. However, in finite systems, some of the particles are frozen. The behavior of finite-sized systems is interesting on its own right  \cite{balogh}, due to the finite extent of numerical simulations \cite{lambda1,lambda2,lambda3,lambda4,adler} but also because of the physical problem of jamming in confined geometries \cite{bi,teomy1,teomy2}. Instead of running the full physical dynamics of these models, a faster way to find $n_{F}$ is to run culling dynamics. In these dynamics the system is scanned iteratively, such that in each step the mobile particles are removed, until either all particles are removed or those remaining cannot be removed. Those that cannot be removed are the frozen particles. For the FA model this process is identical to finding the size of the percolating cluster in bootstrap percolation, in which a site becomes infected (i.e. its state changes to $0$) if at least two of its neighbors are also in state $0$. For the KA model, it is similar with the added requirement that at least one of the neighboring $0$'s have another neighboring $0$. 

By varying the vacancy density $v$ and fixing the system size $L\times L$, the average fraction of frozen particles changes from $n_{F}=0$ at $v=1$ to $n_{F}=1$ at $v=0$. For large systems the transition between $n_{F}=0$ and $1$ occurs over a very narrow range of densities. The critical vacancy density, $v_{c}(L)$, is defined as the density at which on average half of the particles are frozen in a system of size $L\times L$. Holroyd proved \cite{holroyd} that for asymptotically large systems, the critical vacancy density below which the system is highly likely to contain frozen particles is
\begin{align}
v_{c}=\frac{\lambda_{0}}{\ln L} ,\label{eqlam}
\end{align}
with
\begin{align}
\lambda_{0}=\frac{\pi^{2}}{18}\approx0.55 .
\end{align}
Because this result is for asymptotically large systems, it is valid for both periodic and hard-wall boundary conditions.
This result is obtained by considering critical droplets, which are small unfrozen regions that expand to unfreeze the entire system. Holroyd showed \cite{holroyd}, and a sketch of the proof is given in Section \ref{critdrop} below, that in the asymptotic limit the probability that a site seeds a droplet is
\begin{align}
P(v)=\exp\left(-2\lambda_{0}/v\right) .
\end{align}
Since a system is unfrozen if it contains such a droplet, Eq. (\ref{eqlam}) is derived by assuming that at the critical density there is on average one critical droplet in the system, i.e. that
\begin{align}
1=P(v_{c})L^{2} .\label{pl2}
\end{align}

Culling simulations for finite-size systems indicate that indeed $v_{c}\approx\lambda/\ln L$. The value of $\lambda$ was estimated by finding the critical vacancy density $v_{c}$ at which the probability that the system is frozen equals $0.5$ \cite{footnote}, and using Eq. (\ref{eqlam}). However, the value of $\lambda=v_{c}\ln L\approx0.25$ estimated by simulations of both the FA and KA models with $L$ up to $10^5$ is very far from its asymptotic limit $\lambda_{0}\approx0.55$ \cite{lambda1,lambda2,lambda3,lambda4,adler}. Holroyd subsequently showed \cite{slowcon,slowcon2} that the convergence to the asymptotic value is slow, and that the correction to $\lambda$ is
\begin{align}
\lambda_{0}-\lambda=f(v_c)>0 ,
\end{align}
such that for small $v_c$
\begin{align}
O\left(\sqrt{v_c}\ln^{3}(1/v_c)\right)\geq f(v_c)\geq O\left(\sqrt{v_c}\right) .\label{fdef}
\end{align}

The simulations of the largest systems we are aware of are for $L=128,000$ and $v_{c}\approx0.023$ \cite{adler}. De Gregorio et. al. \cite{modified} circumvented the need to simulate large systems in a related model, the modified bootstrap percolation model, by explicitly calculating $P(L,v)$, i.e. the probability that a square of size $L\times L$ is emptied by a single seed. These calculations are for equivalent systems of up to size $L=220,000$. Formally, the same approach can be used for the FA and KA models, but in practice the calculations become too cumbersome.

In this paper we present a new technique that also circumvents the need to simulate large systems, and the results we obtain are for an equivalent system of size $L>10^7$ with $v_{c}=0.016$. In our approach we fix the vacancy density, $v$, and numerically calculate $P(\infty,v)$. We do this by expanding critical droplets and considering only the sites in their vicinity. In effect, we treat the expansion process as a Markov process. In this way there is no need to generate large configurations or irrelevant data. Our results for $f(v)$ are consistent with the theoretical bounds, Eq. (\ref{fdef}). Moreover, we find that the average number of sites seeding a critical droplet is not $1$ but higher at about $6$ for the FA model and $4$ for the KA model at the critical density, and find the corresponding correction to Eq. (\ref{pl2}).

In Section \ref{critdrop} we sketch the derivation of Eq. (\ref{eqlam}). In Section \ref{secseeds} we show numerically that the average number of seeding sites is larger than $1$. In Section \ref{secplv} we show our results for $P(L,v)$ and for $\lambda$. Section \ref{secalgorithm} describes our algorithm, and Section \ref{secsummary} summarizes the paper.

\section{Critical Droplets}
\label{critdrop}

The main idea behind Holroyd's proofs is the notion of critical droplets, which are small unfrozen regions that expand to unfreeze the entire system. Since the droplet can only expand, this method is equivalent to the culling dynamics described above, and not to the physical dynamics. Assuming that the system is either completely frozen or completely unfrozen, the critical density may be defined as the density at which the system contains on average one critical droplet, and so $v_{c}$ is found from Eq. (\ref{pl2}). 

In order to find an analytical approximation for $P(v)$, consider an emptiable square of size $\ell\times\ell$ and randomly choose one of its four sides. With probability $1-\rho^{\ell}$ (where $\rho=1-v$ is the particle density) at least one of the sites adjacent to the square in the chosen direction is empty, and thus in the $m=2$ FA and KA models this square may be expanded to an emptiable rectangle of size $(\ell+1)\times\ell$. With probability $\rho^{\ell}$ all the sites in this direction are occupied. In this case, we check the row or column adjacent to that fully occupied row or column. With probability $1-\rho^{\ell}$ at least one of the sites in that row or column is empty, and thus the $(\ell+2)\times\ell$ rectangle is emptiable. If the second row or column is fully occupied as well, we stop the expansion process. From the emptiable rectangle of size $(\ell+k)\times\ell$ we randomly choose one of its two longer sides, and repeat the same check for the two rows or columns adjacent to the chosen side, and stop the process only if both of them are full. Since we always try to expand the rectangle from its long side, we only need to consider $k\leq2$. Hence, there are in total six states to consider.

We denote $P\left[(\ell+k)\times\ell,v\right]$ as the probability that during the expansion process the rectangle of size $(\ell+k)\times\ell$ is emptiable, with $k\geq0$, and $P_{1}\left[(\ell+k)\times\ell,v\right]$ as the probability that the $(\ell+k)\times\ell$ rectangle is emptiable but the adjacent row is fully occupied. Therefore we have in total six states for each $\ell$, because $k=0,1,2$. Since we know the probability to change from each state to another, we can write a recursion relation relating the probability to reach each of the six states given the probability to reach a smaller rectangle
\begin{widetext}
\begin{subequations}
\begin{alignat}{6}
&P\left[\ell\times\ell,v\right]=\left(1-\rho^{\ell}\right)P\left[\ell\times(\ell-1),v\right]+\left(1-\rho^{\ell}\right)P_{1}\left[\ell\times(\ell-2),v\right] ,\\
&P\left[(\ell+1)\times\ell,v\right]=\left(1-\rho^{\ell}\right)P\left[\ell\times\ell,v\right]+\left(1-\rho^{\ell+1}\right)P\left[(\ell+1)\times(\ell-1),v\right]+\left(1-\rho^{\ell}\right)P_{1}\left[\ell\times(\ell-1),v\right] ,\\
&P\left[(\ell+2)\times\ell,v\right]=\left(1-\rho^{\ell}\right)P_{1}\left[\ell\times\ell,v\right] ,\\
&P_{1}\left[\ell\times\ell,v\right]=\rho^{\ell}P\left[\ell\times\ell,v\right] ,\\
&P_{1}\left[(\ell+1)\times\ell,v\right]=\rho^{\ell+1}P\left[(\ell+1)\times\ell,v\right] ,\\
&P_{1}\left[(\ell+2)\times\ell,v\right]=\rho^{\ell+2}P\left[(\ell+2)\times\ell,v\right] .
\end{alignat}
\end{subequations}
Solving Eqs. (7c)-(7f), and using the solution in (7a)-(7b) yields
\begin{subequations}
\begin{alignat}{2}
&P\left[\ell\times\ell,v\right]=\left(1-\rho^{\ell}\right)P\left[\ell\times(\ell-1),v\right]+\left(1-\rho^{\ell}\right)\left(1-\rho^{\ell-2}\right)\rho^{2\ell-2}P\left[(\ell-2)\times(\ell-2),v\right] ,\\
&P\left[(\ell+1)\times\ell,v\right]=\left(1-\rho^{\ell}\right)P\left[\ell\times\ell,v\right]+\left(1-\rho^{\ell+1}\right)\left(1-\rho^{\ell-1}\right)\rho^{\ell-1}P\left[(\ell-1)\times(\ell-1),v\right]+\nonumber\\
&+\left(1-\rho^{\ell}\right)\rho^{\ell}P\left[\ell\times(\ell-1),v\right] .
\end{alignat}
\end{subequations}
Solving Eq. (8a) for $P\left[\ell\times(\ell-1),v\right]$, and using the result in Eq. (8b) yields
\begin{align}
&P\left[(\ell+1)\times(\ell+1),v\right]=\left(1-\rho^{\ell+1}\right)\left\{P\left[\ell\times\ell,v\right]+\left(1-\rho^{\ell-1}\right)\rho^{\ell-1}P\left[(\ell-1)\times(\ell-1),v\right]-\right.\nonumber\\
&\left.-\left(1-\rho^{\ell}\right)\left(1-\rho^{\ell-2}\right)\rho^{3\ell-2}P\left[(\ell-2)\times(\ell-2),v\right]\right\} .
\end{align}
\end{widetext}
Assuming a solution of the form
\begin{align}
P\left[\ell\times\ell,v\right]=\prod^{\ell}_{k=1}\beta\left(\rho^{k}\right) ,
\end{align}
and further assuming that for large $\ell$ and small $v$ the function $\beta$ depends only weakly on $\ell$, we have a cubic equation on $\beta\left(\rho^{\ell}\right)$
\begin{align}
\beta^{3}(x)=\left(1-x\right)\left[\beta^{2}(x)+\left(1-x\right)x\beta(x)-\left(1-x\right)^{2}x^{3}\right] ,
\end{align}
with the three solutions
\begin{align}
&\beta_{0}(x)=-x\left(1-x\right) ,\nonumber\\
&\beta_{\pm}(x)=\frac{1-x^{2}\pm\sqrt{\left(1-x\right)^{3}\left(1+3x\right)}}{2} .
\end{align}
For asymptotically large squares $(\ell\rightarrow\infty)$ we have $x\rightarrow0$, i.e. the only non-zero solution is $\beta_{+}$ which converges to $1$. Therefore, for large $\ell$ we may approximate $P\left[\ell\times\ell,v\right]$ as
\begin{align}
&P\left[\ell\times\ell,v\right]\approx\prod^{\ell}_{k=1}\beta_{+}\left(\rho^{k}\right)=\exp\left[\sum^{\ell}_{k=1}\ln\beta_{+}\left(\rho^{k}\right)\right] .
\end{align}
Changing the sum over $k$ to an integral over $x=\rho^{k}$ and taking the limit $\ell\rightarrow\infty$, yields
\begin{align}
&P\left[\infty\times\infty,v\right]=\exp\left(-2\lambda_{0}/v\right) ,
\end{align}
with
\begin{align}
\lambda_{0}=-\int^{1}_{0}\frac{\ln\beta_{+}(x)}{x}dx=\frac{\pi^{2}}{18}\approx0.55 .
\end{align}

This asymptotic value of $\lambda_{0}$ was derived by considering only the most likely way to expand the droplet to the shape of a square, since we stopped the expansion if two adjacent rows or columns are fully occupied. By considering more ways to expand the droplet, but still not all of them, Holroyd derived the bounds on $\lambda$ given in Eq. (\ref{fdef}) above.

\section{Number of Seeding Sites}
\label{secseeds}

\begin{figure}
\includegraphics[width=200pt,height=200pt]{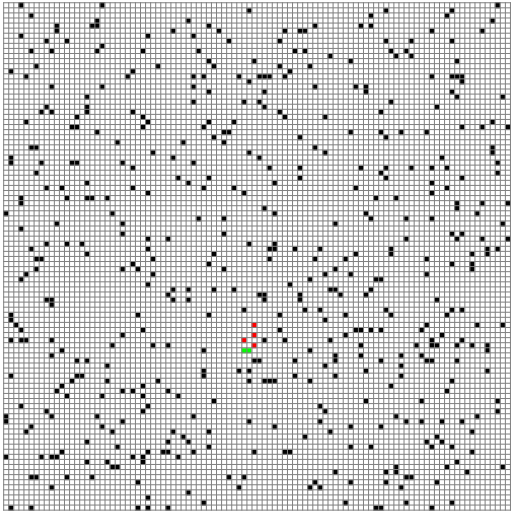}
\caption{A typical configuration for an unfrozen $100\times100$ system at the critical vacancy density in the FA model, $v=0.051$. The white sites are occupied, the black sites are vacancies, the two green (light gray) sites are vacancies that seed a critical droplet according to both the FA and KA rules, and the four red (dark gray) sites are vacancies that seed a droplet only according to the FA rules.}
\label{conf}
\end{figure}

\begin{figure}
\includegraphics[width=\columnwidth]{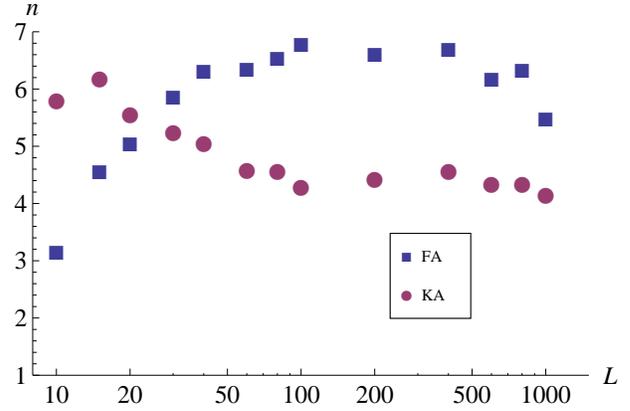}
\caption{The average number of seeding sites in unfrozen configurations, $n$, as a function of the system's size $L$, near the critical density for the FA and KA models. It appears that as $L$ increases, the average number of seeds may be decreasing. However, at such large values of $L$, the value of $n$ is no longer important for the calculation of $\lambda(v)$. The fluctuations exist due to the proximity to the critical density.}
\label{avseeds}
\end{figure}

The customary way to extract $\lambda(v)$ numerically is to say that at the critical density, when half of the configurations in the ensemble are frozen and half are unfrozen, a square of size $L\times L$ has on average one site that seeds a critical droplet.
However, there are correlations between the seeding sites, i.e. if one site seeds a droplet at least another site can also seed a droplet, hence the average number of seeding sites at the critical density is not necessarily $1$, see Fig. \ref{conf} which shows a typical unfrozen configuration.
By defining $n$ as the average number of seeding sites in configurations that have seeds at the critical density, we find that Eq. (\ref{pl2}) should be modified to
\begin{align}
\frac{n}{2}=L^{2}P(v) ,
\end{align}
or equivalently
\begin{align}
\lambda(v)=v\ln\left(\frac{L}{\sqrt{n/2}}\right) .
\end{align}
For very large $L$, when $\ln L\gg\ln \sqrt{n/2}$, the value of $n$ has a negligible effect on the value of $\lambda(v)$. As shown below, we find that for $L\leq1000$ the value of $n$ is $\approx4-7$, and decreases with increasing $L$. For $L>1000$ and $n\leq7$ we find that $\ln L> 10\ln\sqrt{n/2}$, and thus the value of $n$ has only a negligible effect on $\lambda(v)$ for systems with $L>1000$.  We numerically find that the average number of seeding sites is not $2$, but higher at about $5-7$ (FA) or $4$ (KA), as shown in Fig. \ref{avseeds}.

\section{Numerical Calculation of the seeding probability}
\label{secplv}

\begin{figure}
\includegraphics[width=\columnwidth]{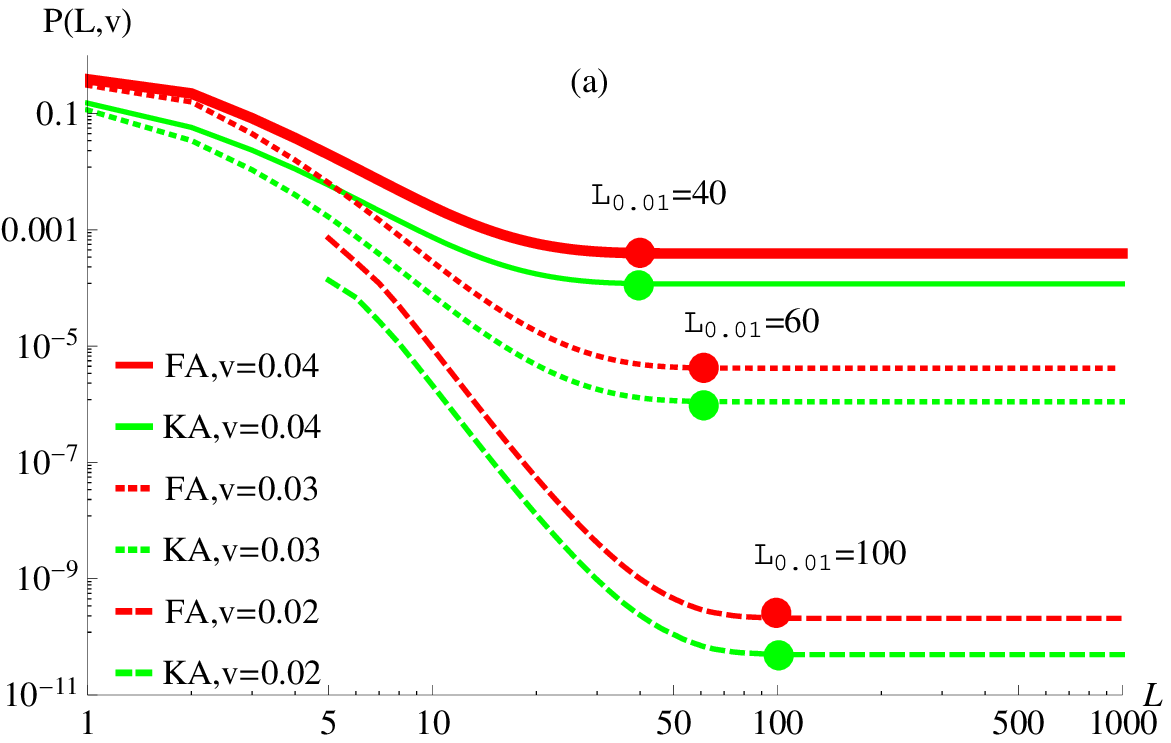}\\
\includegraphics[width=0.5\columnwidth]{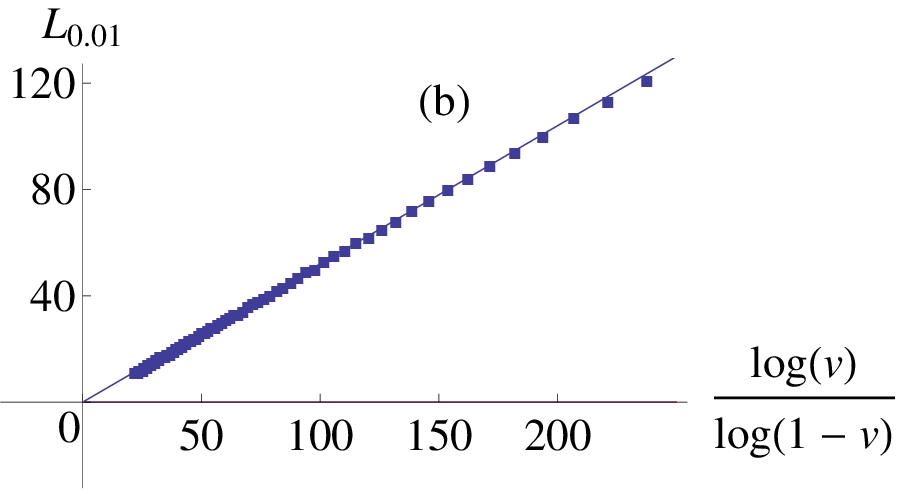}
\includegraphics[width=0.4\columnwidth]{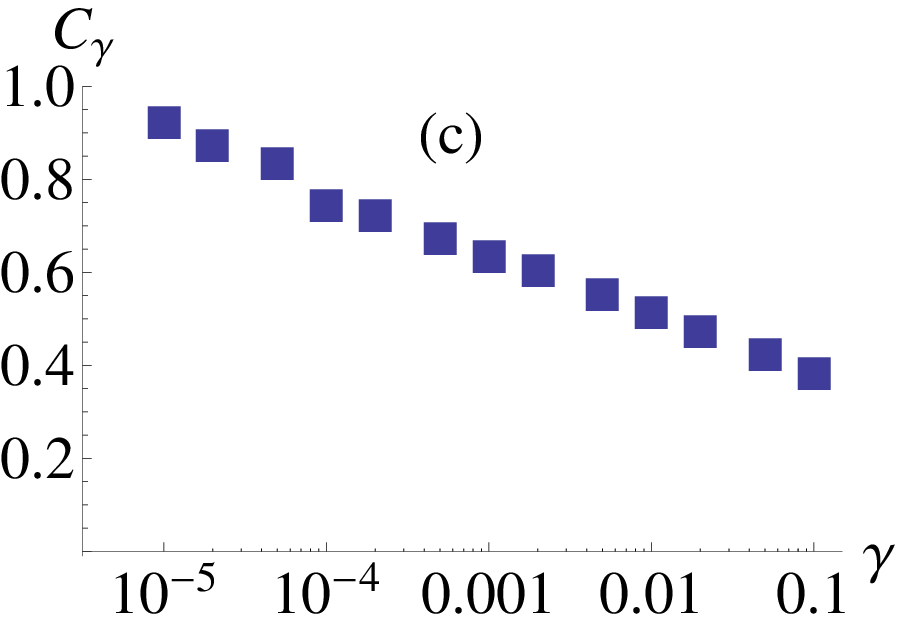}
\caption{(a) The probability that a site seeds a critical droplet of size $L\times L$, $P(L,v)$, as a function of $L$, for $v=0.02,0.03$ and $0.04$. The large dots mark the saturation length $L_{0.01}$. These densities are the critical densities for $L\approx600$ ($v=0.04$), $L\approx6000$ ($v=0.03$), and $L\approx7\times10^5$ ($v=0.02$), far above the saturation length, $L_{0.01}$. The length corresponding to $v=0.02, L=7\times10^5$ is only an estimate because no one ever performed simulations on such large systems. (b) The beginning of the plateau, $L_{0.01}$, as a function of $\ln(v)/\ln(1-v)$. The dots are the numerical results, and the continuous line is $L_{0.01}=C_{0.01}\ln v/\ln(1-v)$, with $C_{0.01}=0.52$. The value of $L_{\gamma}$ for the FA and KA model is exactly the same for all values of $v$ and $\gamma$ we checked. (c) The value of the prefactor $C_{\gamma}$ vs. the choice of the threshold $\gamma$.}
\label{pvconv}
\end{figure}

The main point of this paper is that an alternative way to calculate $\lambda(v)$ is by explicitly calculating the probability to seed a critical droplet. We start from a single site, and check numerically what is the probability $P(L,v)$ that it seeds a critical droplet that unfreezes a region of size $L\times L$. We see from Fig. \ref{pvconv}a that for a given vacancy density $v$, the probability $P(L,v)$ converges rapidly to $P(\infty,v)\equiv P(v)$.
Therefore, from the value $P(L,v)$ at the plateau we can find $\lambda(v)$.

We arbitrarily choose a threshold $\gamma=0.01$ and define $L_{\gamma}$ as the beginning of the plateau, i.e. as the $L$ for which $\left[P(L,v)-P(L_{max},v)\right]/P(L_{max},v)=\gamma$, where $L_{max}=1000$ is the maximal size we consider in our simulations. 
From Fig. \ref{pvconv}b we see that $L_{0.01}$ is approximately given by
\begin{align}
L_{0.01}\approx C_{0.01}\frac{\ln v}{\ln(1-v)} ,
\end{align}
i.e. it scales as the solution to the equation $v=\left(1-v\right)^{L}$. The prefactor of $C_{0.01}=0.52$ is due to the choice of $\gamma=0.01$ as the threshold. If the threshold is taken to $0$, the prefactor goes to $C_{0}=1$ (see Fig. \ref{pvconv}c), but then there are numerical fluctuations at low $v$. The reason for the scaling of $L_{\gamma}$ is due to the different ways in which the droplet may be expanded. A droplet of size $\ell_{1}\times\ell_{2}$ may be expanded to a droplet of size $\ell_{1}\times(\ell_{2}+1)$ if at least one of the sites on the adjacent row of length $\ell_{1}$ is vacant, and it may be expanded to a droplet of size $(\ell_{1}+1)\times(\ell_{2}+1)$ if the site adjacent diagonally to its corner is vacant. The former process has a probability of $1-\rho^{\ell_{1}}$, while the latter has a probability of $v$. At small $L$ the probability that the droplet expands by the first process is very small, while at large $L$ the probability is almost unity. Therefore, once the droplet reaches a critical size, it is highly likely to continue expanding to infinity.

\begin{figure}
\includegraphics[width=1.1\columnwidth]{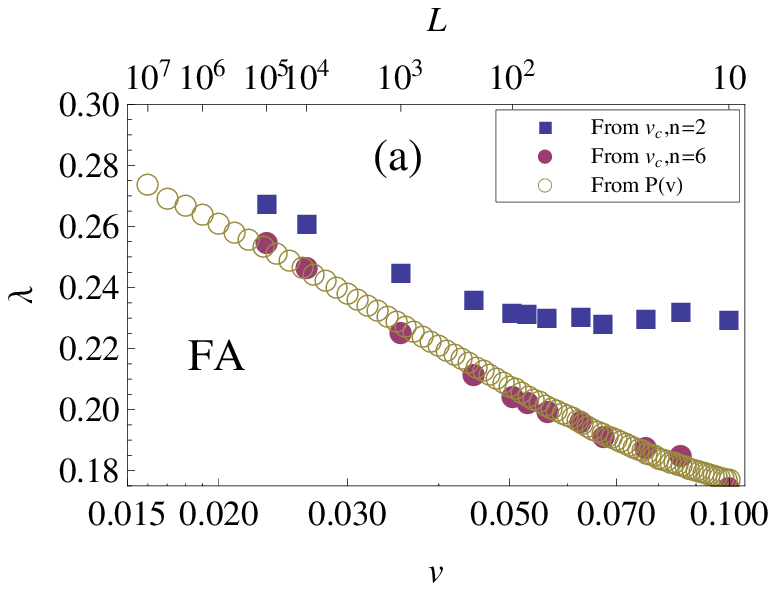}\\
\includegraphics[width=1.1\columnwidth]{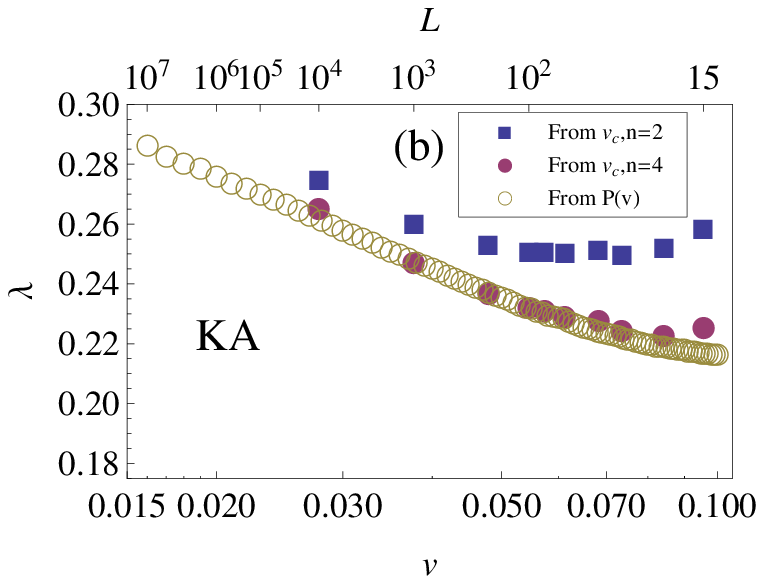}
\caption{The value of $\lambda(v)$ calculated in three different ways for the FA model (a) and the KA model (b): from the critical density of a square of size $L\times L$ with $n=2$, from the critical density of a square of size $L\times L$ with $n=6$ (FA) or $n=4$ (KA), and from $P(v)$. The last two methods give the same results. The values of $L$ at the top horizontal axis correspond to the size of the simulated square.}
\label{lambda}
\end{figure}

The values of $\lambda(v)$ calculated by our method and the traditional method are shown in Fig. \ref{lambda}, where for the traditional method we used $n=6$ for the FA model and $n=4$ for the KA model, see Fig. \ref{avseeds}.
Since the two methods yield the same results, we can say that the corrections to $\lambda$ are not finite-size effects, since these do not affect our method, but rather finite-density effects.

The smallest vacancy density we simulated is $v=0.016$, which corresponds to $\lambda=0.274$ ($0.286$) and squares of size $L=3\times10^7$ ($6\times10^7$) for the FA (KA) model. The running time for $v=0.016$ was $8$-cpu-years, or one month of 96 jobs in parallel, compared to $1.5$-cpu-years for $v=0.017$. As an estimate, we could in principle simulate also $v=0.015$ by enslaving all our computing resources for several months.

\begin{figure}
\includegraphics[width=0.6\columnwidth]{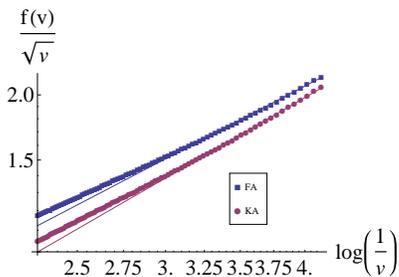}
\caption{Log-log plot of $f(v)/\sqrt{v}$ vs. $\ln(1/v)$. The lines are the fits for $f(v)=A\sqrt{v}\ln^{\alpha}(1/v)$.}
\label{ffit}
\end{figure}

Based on the bounds given in Eq. (\ref{fdef}), we fitted the results of $f(v)=\lambda_{0}-\lambda(v)$, excluding the largest values of $v$, to a function of the form
\begin{align}
f(v)=A\sqrt{v}\ln^{\alpha}(1/v) .
\end{align}
$\alpha=0$ would correspond to the lower bound, and $\alpha=3$ to the upper bound.
We found that $A_{FA}\approx0.44$, $A_{KA}\approx0.36$, $\alpha_{FA}\approx1.1$, and $\alpha_{KA}\approx1.2$, as shown in Fig. \ref{ffit}. Since $0\leq\alpha\leq3$, this functional form is consistent with Eq. (\ref{fdef}).
Using these fits, we can extrapolate to check what system size is needed for $\lambda(v)$ to be, for example, $99\%$ of $\lambda_{0}$. The vacancy density at which this occurs is $v\approx3.7\times10^{-7}$ (FA) or $v\approx2.9\times10^{-7}$ (KA), which corresponds to the critical density of a square of size $L\approx10^{6\times10^5}$ (FA) or $10^{8\times10^5}$ (KA), much larger than any physical system (for comparison, the size of the observable universe is $10^{64}$ times the planck length). For $L=10^{24}$, the critical density is $v_{c}\approx0.006$, which corresponds to $\lambda(v)\approx0.33$ in both models. The vacancy densities and the corresponding sizes needed for $\lambda(v)$ to be a certain percent of $\lambda_{0}$ are shown in Table \ref{tablevl}.

\begin{table}
\begin{tabular}{c|c|c|c|c}
$\frac{\lambda(v)}{\lambda_{0}}$&$v(FA)$&$v(KA)$&$L(FA)$&$L(KA)$\\\hline
$0.6$&$7.6\times10^{-3}$&$9.0\times10^{-3}$&$10^{18}$&$10^{15}$\\
$0.7$&$2.9\times10^{-3}$&$3.1\times10^{-3}$&$10^{58}$&$10^{53}$\\
$0.8$&$8.4\times10^{-4}$&$8.5\times10^{-4}$&$10^{230}$&$10^{220}$\\
$0.9$&$1.2\times10^{-4}$&$1.2\times10^{-4}$&$10^{1700}$&$10^{1800}$\\
$0.95$&$2.1\times10^{-5}$&$1.9\times10^{-5}$&$10^{1.1\times10^{4}}$&$10^{1.2\times10^{4}}$\\
$0.99$&$3.7\times10^{-7}$&$2.9\times10^{-7}$&$10^{6\times10^5}$&$10^{8\times10^5}$
\end{tabular}
\caption{The values of the vacancy density and system size needed for $\lambda(v)$ to be close to $\lambda_{0}$.}
\label{tablevl}
\end{table}

\section{Simulation Algorithm}
\label{secalgorithm}
In our simulations we use a special property of the FA and KA models: an unfrozen region must be a rectangle. This is also true in higher dimensions, but only if $m=2$ vacant neighbors are needed to empty a site. This special property does not exist if $m\geq3$ neighbors are needed.

We start from a single empty site at the origin. Now, consider its neighbor at position $(0,1)$, which is empty with probability $v$. If it is empty, the empty region now contains the origin and the additional site, and so is a rectangle of size $1\times2$. If the site is not empty, consider the site $(0,2)$, which is also empty with probability $v$. If this site is empty, then the site $(0,1)$ has two neighboring vacancies, and so can also be emptied. Hence, in this case the empty region is a rectangle of size $1\times3$. We continue along this direction until we have an empty region of size $1\times k$ and the two sites at $(0,k+1)$ and $(0,k+2)$ are not empty. Now, we consider the sites to the right of the empty rectangle. If at least one of the $k$ sites is empty, with probability $1-\left(1-v\right)^{k}$, the empty region is expanded to a rectangle of size $2\times k$. We now continue in this manner and check the sites surrounding the empty rectangle on all sides. We also remember how many of the adjacent sites to the empty regions we already checked and found that they are occupied.

The main advantage of this algorithm is the negligible memory required, which does not grow as the rectangle grows, since the only variables we need to keep in memory are the size of the rectangle and the number of occupied sites adjacent to it in each of its directions. The history of the expansion process and the exact location of the occupied sites are irrelevant to the final result. Also, since the output of the algorithm is the number of times it was able to unfreeze a large system, it can be easily parallelized, with the final output being the sum of the outputs of the individual jobs.

Extending this algorithm to higher dimensional models with $m=2$ is straightforward, since the unfrozen region must be a hyper-rhomboid, such that at any stage in the algorithm the only variables which should be kept in memory are the size of the hyper-rhomboid ($d$ variables, with $d$ being the dimension of the system), the number of occupied checked sites in the two layers adjacent to each of the $2d$ sides ($4d$ variables), and whether the corners are occupied or not ($2^d$ variables), for a total of $2^d+5d$ variables. For $m\geq3$, the extension of the algorithm is not trivial, since the unfrozen region does not have to be a hyper-rhomboid. In this case, a single droplet can still be expanded, but the memory required is higher since its structure is more complicated.

\section{Summary}
\label{secsummary}
By numerically calculating the probability $P(L,v)$ to expand droplets in the FA and KA kinetically-constrained models we showed that even for rather small $L$, $\lambda(L,v)$ does not depend on $L$, which means that the difference between its value for finite $L$ and $v$ and between its asymptotic value is a finite-density effect, and not a finite-size effect. Our numerical results for $\lambda(v)$ are consistent with the known theoretical bounds. The results we obtained are for an equivalent system of size $L>10^7$, much larger than any previously simulated $(L\approx10^5)$.

Using the data for $\lambda(v)$, we may now check numerically previous results derived for the FA and KA models. For example, Toninelli proved \cite{toniphd} that for asymptotically small $v$ in an infinitely large system in the KA model, the diffusion coefficient of the particles is $D=\exp\left(-2\lambda_{0}/v\right)$. It would be interesting to check whether for a finite density, the diffusion coefficient behaves in the same way with $\lambda(v)$.

The concept behind the algorithm we propose here may be implemented in other models and for other purposes. The idea is to use the fact that the expansion process is Markovian, and to generate only the needed local information without generating in advance unnecessary data. Consider for example the problem of first passage for a tracer particle diffusing in a dense environment \cite{benichou1,benichou2}: how long does it take a particle to move a certain distance from its initial position? At short times, there will be no effect from particles which are far from the tracer particle, so there is no need to follow them or even generate them until they become relevant. The algorithm may be implemented in the following way using the Monte-Carlo method: At time $t=0$, generate the tracked particle at the origin. Since there is at this time only one known particle, and each particle on average attempts to move once every time unit, first advance the clock by one time unit. Now, generate the particles interacting with the tracked particle (assuming the interactions are short ranged) and move the tracer particle in the randomly chosen direction according to its interaction with the surrounding particles. Assuming that in the previous step $N$ new particles were generated, advance the clock by $1/(1+N)$ and randomly choose one of the $N+1$ particles. Generate any new particles which may interact with the chosen particle and move them. Since the interactions are short-ranged, new particles will be generated at each step only if the chosen particle is at the edge of the generated system. In this way, an effectively infinite system may be simulated without using periodic boundary conditions.

\section*{Acknowledgements}

We thank Roman Golkov and Haim Diamant for helpful discussions. This research was supported by the Israel Science Foundation grants No. $617/12$, $1730/12$.


\begin{thebibliography}{999}
\bibitem{footnote}{At the critical density, half of the systems are frozen, because each system in itself is either almost completely frozen or completely unfrozen.}
\end{thebibliography}
\end{document}